\documentclass[12pt]{article}
\usepackage{amsfonts,amsmath,amssymb}
\usepackage{scalerel}[2016/12/29]
\usepackage{graphicx}
\usepackage{tikz}

\newcommand{\2}{\scaleobj{0.8}{\sqrt{2}}}

\def\be{\begin{equation}}
\def\ee{\end{equation}}
\def\bea{\begin{eqnarray}}
\def\eea{\end{eqnarray}}

\def\half{{\textstyle {1\over 2}}}

\begin{document}

\thispagestyle{plain}

\title{\bf\Large Mapping$~$the$~$Calogero$~$model$~$on$~$the$ ~$Anyon$~$model}

\author{St\'ephane Ouvry$^*$  {\scaleobj{0.9}{\rm and}}  Alexios P. Polychronakos$^\dagger$}


\maketitle

\begin{abstract}
We explicitly map the $N$-body one dimensional Calogero eigenstates in a harmonic well to the lowest Landau level sector of $N$-body eigenstates of the two dimensional anyon model in a harmonic well. The mapping is achieved in terms of a convolution kernel that uses as input the scattering eigenstates of the free Calogero model on the infinite line, which are obtained in an operator formulation.
\end{abstract}

\noindent
* LPTMS, CNRS,  Universit\'e Paris-Sud, Universit\'e Paris-Saclay,\\ \indent 91405 Orsay Cedex, France; {\it stephane.ouvry@u-psud.fr}

\noindent
$\dagger$ Department of Physics, City College of New York, NY 10038, USA; \\ \indent
{\it apolychronakos@ccny.cuny.edu}

\section{ Introduction }
Quantum statistics play a crucial role in our understanding of nature. Since the late seventies physicists realized that  non standard statistics, beyond the usual Bose and Fermi cases,   do exist in lower dimensions.  In 2d one speaks of exchange/anyon statistics \cite{LM}, in 1d of fractional/exclusion statistics \cite{alexios},\cite{Haldane}. A paradigmatic experimental setting where these statistics  could play a role is the Quantum Hall effect, where a 2d electron gas is coupled to a strong magnetic field at low temperature. 
Indeed the robustness of quantum Hall states and their properties in the presence of otherwise “messy” interactions is a striking physical effect. A fruitful angle of approach that at least partially accounts for this robustness is the identification of fractional quantum Hall (FQH) excitations as particles obeying anyonic statistics in two spatial dimensions. Nevertheless, and in spite of the intense theoretical study they have received, a complete description of such states is still an open issue.

This work aims at elucidating certain still ill understood aspects of  quantum models with nontrivial statistics, in particular the relation between the anyon and the Calogero integrable models.  

Integrable systems of particles and spin chains have been the object of fascination in physics and mathematics for the last several decades \cite{Calogero}-\cite{OlshaPer}. (For reviews relevant to the present context see \cite{rev}.) Relations between these systems and anyons started emerging since many-body systems of the Calogero type were interpreted as particles of generalized (fractional) statistics in one spatial dimension \cite{alexios}. The connection becomes fully relevant upon the realization that FQH states are generically lowest Landau level (LLL) states (or low-level Landau states) and through a phase space reduction they become effectively one dimensional (although a careful limit needs to be taken to fully capture the one-dimensional thermodynamics \cite{ouvry}). Further, the similarity between Calogero wavefunctions and anyon LLL wavefunctions \cite{polan} suggests a complete mapping between these states in which the real one-dimensional coordinate of Calogero particles maps to a complex coordinate on the plane. This mapping, however, is quite nontrivial, involving either alternative realizations of operators \cite{BH} 
or a matrix model noncommutative realization of FQH states \cite{alexiosbis}.

The intriguing connection between these systems calls for a deeper and more complete mapping of their states, including density correlations. The fact that calculations of such quantities in either system are challenging offers additional motivation to establish their connection as an analytical device whenever one of the systems offers a better setting than the other. 

We are going to explicitly show that the correspondence alluded to above  for the eigenenergies  and the thermodynamics of the LLL-anyon and Calogero models \cite{ouvry} also holds for the eigenstates; more precisely, that a special class of $N$-body anyonic eigenstates in a 2d  harmonic well --the so-called ``linear states"-- coincide with  the $N$-body 1d Calogero eigenstates in a  1d harmonic well. This special class  originates from the LLL anyon  eigenstates    which interpolate  continuously between the complete  LLL-Bose and LLL-Fermi bases \cite{ouvrybis}.  We will demonstrate that an $N$-body kernel exists allowing for a mapping  between the two sets of eigenstates and we will explicitly construct this kernel. Although some technical aspects of our proofs require an integer Calogero coupling constant, we expect our
results to hold for arbitrary values.

\section{The LLL-anyon model}

We consider nonrelativistic anyons of unit mass on the plane with anyonic statistical parameter $\alpha\in[0,1]$ in the presence of an external confining harmonic trap of frequency $\omega$. The noninteracting Hamiltonian for $N$ such anyons is  
\be
\label{start} H_{\rm free}=-2\sum_{i=1}^N\partial_i
\bar\partial_i
+ \sum_{i=1}^N{\omega^2\over 2}\bar z_i z_i
\ee
Anyon statistics is encoded in the monodromy properties of the wavefunction. 
Redefining the $N$-body eigenstate to display explicitly the anyonic monodromy, short-distance behavior and 
long-distance harmonic confinement damping
 \be 
\nonumber\psi_{\rm free}=\prod_{k<l}(z_k-z_l)^{\alpha}e^{- \omega\sum_{i=1}^N z_i\bar z_i/2}\, \psi
\ee
with $\psi$ a bosonic $N$-body state, we obtain a Hamiltonian acting on $\psi$
\bea\label{5}
{H_{\alpha}} &=& -2  \sum _{i=1}^{N} 
              \left[ \partial_i\bar\partial_i 
                -{\omega\over 2}\bar z_i\bar \partial_i
		-{\omega\over 2} z_i \partial_i
		\right]
 \\\nonumber
& & -2\alpha\sum_{i<j}\left[{1\over  z_i-
z_j}{(\bar \partial_i-\bar \partial_j})
-{\omega\over 2}\right]+N\omega
\eea
The Hamiltonian (\ref{5}) acts
trivially on $N$-body eigenstates made of symmetrized products $\prod_{i=1}^N z_i^{\ell_i}$ with $0\le\ell_1\le\ldots \le \ell_N $ of 
the (unnormalized) 1-body 
eigenstates $z_i^{\ell_i}$
so that 
 \be\label{15}
 \psi_{\rm free}=\prod_{i<j} (z_i-z_j)^{\alpha} e^{- \omega\sum_{i=1}^N z_i \bar z_i/2} 
\sum_{\pi \in S_N} \prod_{i=1}^N z_{\pi(i)}^{\ell_i}
\ee
is an eigenstate of the Hamiltonian (\ref{start})
with
an $N$-anyon spectrum
\be\label{16} E_N=\omega\left[
\sum_{i=1}^N
\ell_i
+{1\over2}N(N-1)\alpha\right]+N\omega
\ee
The anyonic eigenstates (\ref{15}) are known in the literature as linear states since their energy (\ref{16}) linearly interpolates in $\alpha$ between bosonic $N$-body harmonic eigenstates for $\alpha=0$  to fermionic ones for $\alpha=1$. Obviousy, they only constitute a small subset of the full $N$-anyon spectrum, which remains so far unknown.

One notes \cite{ouvry} that  
the one body harmonic eigenstates  at hand in (\ref{15}), namely $z_i \;e^{- \omega z_i \bar z_i/2} $, correspond to picking on each ($\ell_i+1$)-degenerate 2d harmonic energy level $\omega(\ell_i+1)$, $\ell_i\ge 0$,
the state of maximal angular momentum $\ell_i$. The resulting one body spectrum is identical to that of an 1d harmonic oscillator 
(apart from a constant  shift) where the 2d orbital quantum number $\ell_i$ now stands for the excitation number associated to 1d one body harmonic eigenstates. Clearly a dimensional reduction has taken place from 2d to 1d.
Furthermore, trading $\alpha$ for $g$, one recognizes in (\ref{16}) the $N$-body 1d Calogero spectrum in a harmonic
well (up to a global $N\omega /2$ shift) with Hamiltonian
\be
{\tilde H}_g= -\frac{1}{2}\sum_{i=1}^N \frac{\partial^2}{\partial x_i^2} + 
\sum_{i<j}\frac{g(g-1)}{(x_i-x_j)^2}
+\frac{1}{2}\,\omega^2\, \sum_{i=1}^N x_i^2
\label{calogero}
\ee
The Calogero model has also been argued to inbue quantum statistics of order $g$ to its particles, and its states carry the
tell-tale $g$-power of the Vandermonde factor. These similarities, and the fact that the two systems have the same energy spectrum and the same degeneracies, makes it clear that there should be a mapping between the two. In the following sections we will indeed establish such an explicit mapping.

Before doing so, let us remark that the eigenstates (\ref{15}) beg  for a physical  interpretation in terms of LLL eigenstates.  Indeed, if we add an external magnetic field  to which  the anyons now couple, we obtain the noninteracting Hamiltonian
\be\label{new}
 H_{\rm free}=-2\sum_{i=1}^N\left[(\partial_i-{\omega_c\over 2}\bar z_i)
(\bar\partial_i+{\omega_c\over 2}z_i) - {\omega_c\over 2}\right]
+ \sum_{i=1}^N{\omega_t^2\over 2}\bar z_i z_i
\ee
where $\omega_c$ is half the cyclotron frequency of the magnetic field  and  $\omega_{\rm t}^2=\omega_{\rm c}^2+\omega^2$.
 Redefining as above the $N$-body eigenstates to display explicitly the anyonic monodromy and long distance harmonic and Landau  damping
 \be 
\nonumber\psi_{\rm free}=\prod_{k<l}(z_k-z_l)^{\alpha}e^{- \omega_t\sum_{i=1}^N z_i\bar z_i/2}\, \psi
\ee we obtain the Hamiltonian acting on $\psi$
\bea\label{5bis}
{H_{\alpha}} &=& -2  \sum _{i=1}^{N} 
              \left[ \partial_i\bar\partial_i 
                -{\omega_t+\omega_c\over 2}\bar z_i\bar \partial_i
		-{\omega_t- \omega_c\over 2} z_i \partial_i
		\right]
\nonumber \\
& & -2\alpha\sum_{i<j}\left[{1\over  z_i-
z_j}{(\bar \partial_i-\bar \partial_j})
-{\omega_t-\omega_c\over 2}\right]+N\omega_t
\eea 
It is obvious that $\prod_i z_i^{\ell_i}$ is still an eigenstate of (\ref{5bis}), and thus (\ref{15}) are eigenstates of (\ref{new}) provided that $\omega$ is replaced by $\omega_t$
with the spectrum  
\be\label{notbad} E_N=(\omega_t-\omega_c)
\left[
\sum_{i=1}^N
\ell_i
+{1\over2}N(N-1)\alpha+N\right]+N\omega_t\ee
In the absence of  the harmonic confinement $\omega=0$, i.e., $\omega_t = \omega_c$,   (\ref{15})  reduces to the  LLL $N$-anyon eigenstates 
\be\label{999}
 \psi_{\rm free}=\prod_{i<j} (z_i-z_j)^{\alpha} e^{-\omega_c\sum_{i=1}^N z_i \bar z_i/2} \;
\sum_{\pi \in S_N} \prod_{i=1}^N z_{\pi(i)}^{\ell_i}
\ee 
 with a degenerate  spectrum $E_N=N\omega_c$  for any statistical parameter $\alpha$. When $\alpha$ varies from 0 to 1,  the LLL-anyon eigenstates (\ref{999})  interpolate continuously between the complete Bose and Fermi LLL states: this is the LLL-anyon model.

Clearly the eigenstates  (\ref{15}) can be viewed as originating from the LLL-anyon eigenstates (\ref{999})  deformed by a harmonic well  in the limit of a vanishing magnetic field. The harmonic confinment lifts the degeneracy and yields the $\alpha$-dependent $N$-anyon spectra (\ref{16}) or (\ref{notbad}). For this reason, from now on we will refer to (\ref{15})  as ``LLL-induced" wavefunctions, and for short  LLL wavefunctions.

\section{The mapping kernel}
Our goal will be to present a convolution kernel that affects an explicit mapping of the wavefunctions of the Calogero
model to the corresponding LLL anyon wavefunctions. That is, we will derive a function of $2N$ variables
$k_g[x,z]$, where $[x,z]$ denotes the collection of position variables $x_1,x_2,\ldots,x_N$ and $z_1,z_2,\ldots,z_N$,
such that
\be\nonumber
\int k_g[x,z] \, \psi_{C,g} [x] \, [dx] = \psi_{a,g} [z]
\ee
where $\psi_{C,g}$ is an eigenstate of the Calogero model in an harmonic well, while $\psi_{a,g}$ is a LLL wavefunction of the
corresponding planar anyon model in terms of holomorphic coordinates $z_i$  (with  $\alpha\to g$ understood).
We will consider a mapping in which the trivial long-distance harmonic dampings $e^{-\omega\sum_i x_i^2/2}$ of $\psi_{C,g}$ and $e^{-\omega\sum_i z_i\bar z_i/2}$ of $\psi_{a,g}$ have been factored out.

We will introduce the method in steps, starting from the relatively simple case of two particles, first free and then
interacting with a Calogero potential, and then proceeding to treat the general case that is substantially more involved. 
Also, from now on we set $\omega=1$.

\subsection{The 2-body kernel}

Let us first examine, for the sake of simplicity, the 2-body problem.
The Calogero Hamiltonian (\ref{calogero}) readily decouples into a center of mass and a relative part. The
center of mass part is a simple harmonic oscillator, independent of $g$, so we focus on the $g$-dependent relative
coordinate part.

First, in the free case $g=0$, 
the relative 2-body eigenstates in a harmonic well in terms of the relative
coordinate $x=x_1 - x_2 $ are 
 \be
\label{1bis}\psi_{n} (x) =  H_{n}(x/{\2})\, e^{-x^2/4} ~,~~~ n =2\ell\ee with energy $n+1/2$
where $H_{n}(x)$ is a Hermite polynomial of even order  and therefore $\psi_{n}$ is by convention bosonic (nontrivial
statistics are expressed through $g$, once $g \neq 0$; $g=1$ is also free but corresponds to fermion statistics).
The Hermite polynomial $H_{n}$ can be obtained as
\be 
\label{hermitebis} 
H_{n} (x/\2) ={\sqrt{2}\,}^n e^{x^2/2}{d^{n}\over dx^{n}} \, e^{-x^2/2}
\ee 
It is easy now to turn (\ref{1bis}) into a 2d eigenstate by means of the 2-body kernel $e^{- (x-iz)^2/2}$. Using (\ref{hermitebis}) and integrating by parts $n$ times we find
\bea
\nonumber\int_{-\infty}^{\infty} e^{-(x-iz)^2/2} H_{n}(x/\2)\, dx \hskip -0.2cm 
&=& \hskip -0.2cm  {\sqrt{2}\,}^n \int_{-\infty}^{\infty} e^{i x z+{}z^2/2}\, {d^{n}\over dx^{n}} e^{-x^2/2}\, dx\\
&=& \nonumber\hskip -0.2cm {\sqrt{2}\,}^n\int_{-\infty}^{\infty} e^{-x^2/2} {d^{n}\over dx^{n}} e^{i x z+z^2/2}\,dx\\
\nonumber
&=& \hskip -0.2cm \left(i\sqrt{2}z\right)^{n}  \int_{-\infty}^{\infty}e^{-x^2/2+i x z +{}z^2/2}\, dx\\\nonumber
&=& \hskip -0.2cm \sqrt{2\pi}\, \Bigl(i\sqrt{2}z\Bigr)^{n}  
\eea
That is,
\be\label{thisis}
\int_{-\infty}^{\infty} e^{-(x-iz)^2/2} H_{n}(x/\2) \, dx 
= \sqrt{2\pi}\, (i\2 z)^{n}   
\ee
i.e., $H_n(x/\2)$ in (\ref{1bis}) is mapped to $z^{n}$ up to an overall coefficient. We recognize the above as the transition from an oscillator basis  to a coherent state basis.

For $g\ne 0$, a similar logic prevails: the relative 2-body Calogero eigenstates in a harmonic well are now 
\be \label{thisistheend}
\psi_{n,g}(x)= x^g \,  H_{n,g}(x/\2)\,e^{-x^2/4} ~,\quad n=2\ell,~~~  x>0
\ee  with energy $n+g+1/2$
and by convention $\psi_{n,g} (-x) = \psi_{n,g} (x)$. The $g$-deformed Hermite polynomials
$H_{2\ell,g}$, providing the polynomial part of the Calogero eigenstates, can be obtained through the action  of generalized ladder operators on the ground state. This will be fully exploited in the next section dealing with the $N$-body case. Here we prefer to take advantage of the relative simplicity of the 2-body case to provide a more direct
and straightforward treatment that will motivate the $N$-body result.

The deformed Hermite polynomials can be obtained as
\be \label{hermiteg} H_{2\ell,g}(x/\2)=2^\ell \, x^{-g} \, e^{x^2/2}\bigg({d^2\over dx^2}-{g(g-1)\over x^2}\bigg)^\ell ( x^ge^{-x^2/2})
\ee 
This arises by acting on the ground state with an even power $2\ell$ of the generalized creation operator for the relative coordinate and
conjugating the result with a gaussian, as in the $g=0$ case. The power $2\ell$ is required to be even by bosonic
statistics.

The operator ${d^2/ dx^2}-g(g-1)/x^2$ that appears above is essentially  the relative 2-body scattering Calogero Hamiltonian on the infinite line, with scattering eigenstates the Bessel functions
\be \nonumber h_{g}(kx)=\sqrt{2\pi}\sqrt{kx} J_{g-1/2}(kx)\ee
which are solutions of
\be \label{hermeq}
\left({d^2\over dx^2}-{g(g-1)\over x^2}\right)h_{g}(kx)=-k^2 \, h_{g}(kx)
\ee
For $k x$ large, $ h_{g}(kx)$ becomes a combination of plane waves  $e^{\pm i k x}$ with momentum $k$ and scattering phase
$e^{-i \pi g}$
\be
\nonumber
h_{g}(kx) \to e^{-i {\pi \over 2} g} \, e^{ikx} + e^{i{\pi \over 2} g} \, e^{-ikx}  \quad ({x \to \infty})
\ee 
while for $k x$ near zero it behaves as $h_g \sim x^g$. Note also that, in the free case $g=0,1$, $ h_{g}(kx)= 2 \cos (kx)$ and
$h_{g}(kx)= 2 \sin (kx)$ respectively.

Let us now show that the appropriate 2-body kernel that will turn Calogero states into anyon states is
\be\nonumber
k_g(x,z)=e^{z^2/2}  h_g(zx) \, e^{-x^2/2} 
\ee
This is the generalization of the kernel for $g=0$, but with the Calogero scattering solution instead of the plane wave $e^{ixz}$.

In the present case we are restricting the convolution
integral to positive values of $x$, since $x^g$ is not well-defined for negative values of $x$ when $g$ is fractional.
Again, by integrating by parts $2\ell$ times, or using the hermiticity of the scattering Calogero Hamiltonian operator in
(\ref{hermeq}), we obtain
\bea \nonumber
&&\hskip -0cm\int_{0}^{\infty} e^{z^2/2}  h_g(xz)\, e^{-x^2/2}~ x^g H_{2\ell,g}(x/\2)\, dx\\ \nonumber
&& \hskip 1.5cm=2^\ell\int_0^{\infty}e^{z^2/2} h_g(xz)\bigg({d^2\over dx^2}-{g(g-1)\over x^2}\bigg)^\ell ( x^ge^{-x^2/2} ) \, dx\\ \nonumber
&&\hskip 1.5cm=2^\ell\int_0^{\infty}e^{z^2/2}x^{g}\,e^{-x^2/2}\bigg({d^2\over dx^2}-{g(g-1)\over x^2}\bigg)^\ell  h_g(xz)\, dx\\
&&\hskip 1.5cm= 2^\ell (-z^2)^{\ell}e^{z^2/2} \int_0^{\infty}x^{g}e^{-x^2/2} h_g(xz)\, dx\label{alexios}
\eea
In the above integrations by parts it was crucial not to have any contribution arising from the boundary terms at $x=0$.
Writing the relation (\ref{hermiteg}) in the recursive form
\be\nonumber
 H_{2\ell,g}(x/\2)=2 x^{-g} e^{x^2/2}\bigg({d^2\over dx^2}-{g(g-1)\over x^2}\bigg) ( x^g e^{-x^2/2} H_{2\ell-2,g}(x/\2))
\ee
we see that the boundary terms picked up in each of the  integrating steps are
\bea \nonumber
&&\hskip -0.7cm\int_{0}^{\infty} h_g(xz)\, e^{-x^2/2}~ x^g H_{2\ell,g}(x/\2)\, dx\\\nonumber
&&\hskip 0cm=2\int_0^{\infty}h_g(xz)\bigg({d^2\over dx^2}-{g(g-1)\over x^2}\bigg)( x^ge^{-x^2/2} H_{2\ell-2,g}(x/\2) )\, dx\\ \nonumber
&&\hskip 0cm=-2z^2 \int_0^{\infty}h_g(xz) \, x^g\, e^{-x^2/2} H_{2\ell-2,g}(x/\2)\,  dx\\ 
&&\hskip 0.47cm - 2\left. h_g(xz){d \over dx}( x^ge^{-x^2/2} H_{2\ell-2,g}(x/\2)) \right|_0 \nonumber
+ 2\left.  x^ge^{-x^2/2} H_{2\ell-2,g}(x/\2) {d \over dx} h_g(xz) \right|_0
\eea
By taking into account the behavior of $h_g (xz) \sim (xz)^g$ near zero and the fact that $H_{2\ell-2}$ is regular and goes to
a constant at zero, we see that the boundary terms cancel. By contrast, had we chosen as $h_g (zx)$ the second solution 
of the Calogero scattering problem that behaves as $x^{1-g}$ near zero, the combined boundary terms would be proportional
to $(1-2g) z^{1-g} \, H_{2\ell-2} (0)$ and would not vanish. 

The last integral $\int_0^{\infty}x^{g}e^{-x^2/2} h_g(xz)\, dx$ in (\ref{alexios}) is an $\ell$-independent function of $z$ that will yield the desired relative 
2-anyon monodromy factor $z^g$, as we will now demonstrate.

When $g$ is an integer, $h_g(x)$ can be rewritten
iteratively as 
\be\nonumber
 h_g(x) = (-1)^g \, 2\, x^g\bigg({1\over x} {d\over dx}\bigg)^g \cos x
\ee
which is essentially Rayleigh's formula for spherical Bessel functions.
It follows that 
\bea
h_g(xz)&=&   (-1)^g \, 2 \, z^{-g}\, x^{g}\bigg({1\over x} {d\over dx}\bigg)^g \cos (xz) \nonumber \\
&=& (-1)^g \, 2\, x^{-g}\, z^{g}\bigg({1\over z} {d\over dz}\bigg)^g \cos (xz) \label{itero}
\eea
where in the last step we used the fact that the function  $h_g(xz)$ is symmetric in $x$ and $z$.
Therefore
\bea\nonumber \int_0^{\infty}x^{g}e^{-x^2/2} h_g(xz) \,dx
&=&(-1)^g \, 2\int_0^{\infty} x^{g}e^{-x^2/2} x^{-g} z^{g}\bigg({1\over z} {d\over dz}\bigg)^g\cos (xz)\, dx\\\nonumber
&=&(-1)^g \, 2 \, z^g\bigg({1\over z} {d\over dz}\bigg)^g\int_0^{\infty}e^{-x^2/2}\cos (xz) dx\\\nonumber
&=&(-1)^g {\sqrt{2\pi}}\, z^g\bigg({1\over z} {d\over dz}\bigg)^g e^{- z^2/2}
\eea
Since ${1\over z} {d\over dz} e^{- z^2/2}=- e^{- z^2/2}$, it follows that $\Bigl({1\over z} {d\over dz}\Bigr)^g e^{- z^2/2}
=(-1)^{-g} e^{ -z^2/2}$. Inserting this in (\ref{alexios}) we finally obtain
\be \nonumber
\int_{0}^{\infty}x^{g}\, e^{-x^2/2}\, h_g(xz)\, dx= {\sqrt{2\pi}}\, z^g\,  e^{-z^2/2} 
\ee
This result was proven for integer values of $g$ in order to introduce a technique similar to the one that will be used in the nontrivial case of arbitrary $N$. However, it actually holds for arbitrary real values of $g$ as it is a known integral property of Bessel functions \cite{Bessel}. Overall, we obtain the desired mapping 
\be\label{oupsoups}
\int_{0}^{\infty} e^{z^2/2} h_g(xz) \, e^{-x^2/2} \,x^g\, H_{2\ell,g}(x/{\2})\, dx\\ 
= (-2)^\ell \sqrt{2\pi} \, z^g \,z^{2\ell}
\ee
from the relative 2-body Calogero eigenstate (\ref{thisistheend})  to the relative 2-anyon eigenstate $ z^g \,z^{2\ell}$, as given in (\ref{15}) for the relative 2-anyon problem  with $\alpha\to g$ understood. Note that for $g=0$ (\ref{oupsoups})  reduces to (\ref{thisis}), since $2\cos(xz) = e^{ixz}+ e^{-ixz}$ reproduces the integral
of $e^{ixz}$ from $-\infty$ to $\infty$.

Reintroducing the center of mass  coordinate $X=(x_1+x_2)/2$, the full energy eigenstates are products of the
above relative Calogero harmonic eigenstates and ordinary ($g=0$) harmonic oscillator eigenstates in $X$. The full mapping functional
is the product of the relative kernel for the set of variables  $x=x_1-x_2$, $z=(z_1-z_2)/2$ and the center of mass kernel  $e^{(2X-i Z/2)^2/2}$ for the set of variables $X=(x_1 + x_2 )/2$, $Z=z_1+z_2$. In terms of the particles variables     $x_1, x_2; z_1, z_2$ this 2-body kernel rewrites as 
\be \label{2-body}
k_g(x_1,x_2;z_1,z_2)=e^{ z_1^2/4 + z_2^2/4}  e^{-x_1^2 - x_2^2} \,
h_g\big(\small{(x_1-x_2)(z_1-z_2)/2}\big)   e^{{i}(x_1+x_2)(z_1+z_2)/2}
\ee
We recognize the factor after the harmonic exponentials as the full energy eigenstate of the scattering 2-body Calogero model
that behaves asymptotically as the plane wave 
$e^{-i{\pi \over 2} g} e^{i(x_1 z_1 + x_2 z_2 )}+e^{i{\pi \over 2} g} e^{i\pi g}\, e^{i(x_1 z_2 + x_2 z_1 )}$.

\subsection{ The $N$-body kernel }

We now aim to generalize the 2-body kernel (\ref{2-body}) to the $N$-body case. The logic is similar, but the method needs to
be substantially generalized. 

The $N$-body ground state of the  harmonic Calogero Hamiltonian  (\ref{calogero}) with $\omega =1$ is
\be \label{gsN}
\psi_0=\prod_{i > j}(x_i-x_j)^g \, e^{- \sum_{i=1}^N x_i^2/2} \,:=\, {\Delta_x}^g \, e^{- [x^2]/2}
\ee
where we defined
\be\nonumber
\Delta_x=\prod_{i >j}(x_i-x_j) ~,~ \quad [x^2] = \sum_{i=1}^N x_i^2
\ee 
$\psi_0$  has energy $E_{g,0}= gN(N-1)/2 + N/2$, i.e., up to a global shift, (\ref{16})  with all the $\ell_i$'s put to zero (with $\alpha\to g$ and $\omega=1$ understood).
As in the 2-body case, the ground state is taken to be
bosonic and (\ref{gsN}) holds for a fixed ordering of the particles, in the ``wedge" $x_1 < \dots < x_N$,
while it is extended symmetrically in the remaining $N!-1$ wedges.

Excited states can be obtained through the action of symmetric products of the ladder operators \cite{polex,bhv}
\be\label{creation} 
a_i^+= \Pi_i - x_i ~, ~~~~~ a_i = -\Pi_i - x_i
\ee
where $\Pi_i$ are the antihermitian operators
\be\nonumber
\Pi_i = {\partial\over \partial x_i}+\sum_{j\ne i} M_{ij}{g \over x_i-x_j}
\ee
with $M_{ij}$ particle permutation operators exchanging particles $i$ and $j$. The operators $\Pi_i$
commute,
\be\nonumber
[ \Pi_i , \Pi_j ] =0
\ee
As will be reviewed in the sequel, energy eigenstates $\psi_{\ell}$, labeled by the $N$ integers $\ell_i, i=1,\dots,N$ 
(where $\ell$ is understood as the set $\ell_1,\ell_2,\ldots,\ell_N$), are obtained by acting on the ground state with 
fully symmetrized monomials of $a_i^+$:
\be 
\psi_{\ell} = \sum_{\pi\in S_N}\prod_{i=1}^N \big( a_{\pi(i)}^+\big)^{\ell_i}\psi_0
\label{stateh}
\ee
We are looking for an $N$-body kernel  $k_g[x,z]$, where $x$ and $z$ denote the collection of variables 
$x_1 < \ldots < x_N$ and $z_1 < \ldots <z_N$  respectively, such that
\be\nonumber
\int  k_g [x,z] (\psi_\ell [x] e^{[x^2]/2}) \, [dx]
={\Delta_z}^g \sum_{\pi\in S_N}  \,
 \prod_{i=1}^N z_{\pi(i)}^{\ell_i}
\ee
that is, $\psi_\ell$ is mapped on the $N$-anyon eigenstate (\ref{15}) (with overall harmonic damping terms removed). The above integral, as well as all other integrals in this section, is understood to be over the fundamental wedge $x_1 < x_2 < \cdots < x_N$.

As in the 2-body case, the kernel will be given in terms of the $N$-body scattering Calogero eigenstates.
The scattering Calogero $N$-body Hamiltonian is
\be\nonumber
{ \bar{H}_{g}}= -\frac{1}{2}\sum_{i=1}^N \frac{\partial^2}{\partial x_i^2} + 
\sum_{i<j}\frac{g(g-1)}{(x_i-x_j)^2}
\ee
i.e., (\ref{calogero}) with $\omega=0$.
We define $h_g [x,z]$ to be the eigenstate of the scattering Calogero model with asymptotic momenta $z_1 < \cdots < z_N$
\be\nonumber
{\bar H}_g \, h_g [x,z] = \half [z^2] \, h_g [x,z]
\ee
that behaves as $(x_i - x_j )^g$ when $x_i$ and  $x_j$ are close and becomes in the asymptotic region the combination of
scattered plane waves
\be \label{asympt}
e^{i \sum_i x_iz_i} \to e^{-i\pi{N(N-1)}g/4} \sum_{\pi\in S_N} e^{i \pi g \, c(\pi)} \, e^{i \sum_i x_{\pi(i)} z_i}
\ee
where $c(\pi)$ is the number of particle crossings needed to achieve $\{x_{\pi (i)}\}$ from $\{ x_i \}$, encoding
the fact that the scattering phase shift is $-\pi g$ per two-particle scattering. The overall phase was chosen to conform
with the $N=2$ case, and it also makes the phases of the incident wave $e^{i( x_1 z_N + \dots + x_N z_1)}$ and the fully
scattered wave $e^{i( x_1 z_1 + \dots + x_N z_N)}$ symmetric.
Then the appropriate kernel is given by
\be\nonumber
k_g [x,z] = e^{[z^2]/4} e^{- [x^2]} \, h_g [x,z]
\ee 
which is the generalization of (\ref{2-body}) to the $N$-body case.
 
The proof proceeds along similar lines as in the 2-body case.
The creation operators (\ref{creation}) can be written as 
\be \label{con}
a_i^+ = e^{ [x^2]/2}\, \Pi_i \, e^{-[x^2]/2}
\ee
The convolution integral over the fundamental wedge $x_1 < \dots < x_N$ then becomes
\bea
&&\hskip -1.5cm \int k_g [x,z] (e^{[x^2]/2} \psi_\ell [x] )\, [dx]\nonumber \\
 &=&\hskip -0.2cm \int e^{[z^2]/4} e^{- [x^2]/2} \, h_g [x,z] \sum_{\pi\in S_N} \prod_{i=1}^N ( a_{\pi(i)}^+ )^{\ell_i} {\Delta_x}^g \, e^{-[x^2]/2}\, [dx] \nonumber\\\nonumber
&=& \hskip -0.1cm\int e^{[z^2]/4}  h_g [x,z] \sum_{\pi\in S_N} \prod_{i=1}^N { \Pi_{\pi(i)}}^{\ell_i} \, {\Delta_x}^g \, e^{-[x^2]}\, [dx] \\
&=& (-1)^{\sum_i \ell_i} e^{[z^2]/4} \int {\Delta_x}^g \, e^{-[x^2]}\sum_{\pi\in S_N} \prod_{i=1}^N  {\Pi_{\pi(i)}}^{\ell_i} h_g [x,z] \,  [dx]\nonumber
\eea
where in the last step we used the antihermiticity of $\Pi_i$, ensured by the fact that the behavior of $h_g [x,z]$ as
$x_i \to x_j$ eliminates boundary terms, as in the 2-body case.

As we will review below, the action of the symmetrized products in $\Pi_i$ on $h_g [x,z]$ in the above expression reduces them to
the conserved integrals of the scattering Calogero model and produces as their eigenvalues the corresponding symmetrized product of $z_i$
\be
\sum_{\pi\in S_N} \prod_{i=1}^N  \Pi_{\pi(i)}^{\ell_i} \,h_g [x,z] = 
\sum_{\pi\in S_N} \prod_{i=1}^N (i z_{\pi(i)})^{\ell_i}  ~h_g [x,z]  \label{eigen}
\ee
This leads to
\be
\int  k_g [x,z] \psi_\ell [x] \, [dx]
= (-i)^{\sum_i \ell_i}\, e^{[z^2]/4}  \sum_{\pi\in S_N} \prod_{i=1}^N  z_{\pi(i)}^{\ell_i} 
 \int {\Delta_x}^g \, e^{-[x^2]} h_g [x,z] \,  [dx] \label{sem}
\ee

We proceed now to justify (\ref{stateh},\ref{eigen}) and to show that the integral in the RHS of (\ref{sem}) above produces the desired
${\Delta_z}^g$ factor. In the process, we shall also obtain an expression of the scattering states of the Calogero model in terms of an operator acting on plane waves. To this end, we define the exchange-Calogero scattering Hamiltonian
\be
\hat{H}_\Pi= -{1\over 2}\sum_{i=1}^N \Pi_i^2 = -\frac{1}{2}\sum_{i=1}^N \frac{\partial^2}{\partial x_i^2} + 
\sum_{i<j}\frac{g(g-M_{ij})}{(x_i-x_j)^2} \label{exsc}
\ee
and its harmonic counterpart
\be\nonumber
\hat{{H}}_a= \sum_{i=1}^N \half (a_i^+ a_i+a_i a_i^+) = \frac{1}{2}\sum_{i=1}^N \Bigl(-\frac{\partial^2}{\partial x_i^2} + x_i^2 \Bigr) + 
\sum_{i<j}\frac{g(g-M_{ij})}{(x_i-x_j)^2}
\ee
The symmetrized products of the commuting operators $\Pi_i$ in (\ref{eigen}) commute among themselves and with the Hamiltonian $\hat{H}_\Pi$ (\ref{exsc}) and represent integrals of motion for $\hat{H}_\Pi$. All the above operators are invariant under particle permutation. The action of exchange-Calogero Hamiltonians on bosonic $N$-body states $\psi_B$ reduces 
to that of regular Calogero Hamiltonians; that is,
\bea\label{hab}
\hat{{H}}_a  \, \psi_B&=&{\tilde{H}}_g \, \psi_B \nonumber \\
\hat{{H}}_\Pi  \, \psi_B&=&{\bar{H}}_g \, \psi_B
\eea
Likewise, the action of the symmetrized products of $\Pi_i$ on bosonic states reduces them to integrals of motion of 
the regular scattering Calogero Hamiltonian. Starting from a (non-symmetric) simultaneous eigenstate of the $\Pi_i$
\be\nonumber
\Pi_i \, {\hat \psi} [x,z] = i z_i \, {\hat \psi} [x,z]
\ee
and projecting it on the bosonic subspace through the projection operator $S = \sum_{\pi \in S} M_\pi$ we obtain the
corresponding simultaneous eigenstate of the Calogero Hamiltonian and all symmetrized products of the $\Pi_i$ (which commute
with the bosonic projection) with eigenvalue the corresponding symmetrized product of $z_i$. This eigenstate must be
$h_g [x,z]$, since it has the same energy and the same asymptotic behavior (the $\Pi_i$ act as 
$\partial / \partial x_i$ when $x_i \to \infty$), which justifies (\ref{eigen}).

Similarly, the harmonic exchange-Calogero Hamiltonian satisfies
\be\label{aa+}
[\hat{{H}}_a , a_i^+]=a_i^+ ~~\quad {\rm and}~~\quad[\hat{{H}}_a , a_i]= -a_i
\ee
Starting with a symmetric state annihilated by all lowering operators $a_i$, which is actually $\psi_0$ in
(\ref{gsN}), and acting with symmetrized products of $a_i^+$ we obtain excited states of $\hat{{H}}_a$. Since these
states are bosonic, they are also eigenstates of the regular harmonic Calogero Hamiltonian, which justifies (\ref{stateh}).

To find the value of the integral in the RHS of (\ref{sem}), we further define the Vandermonde-like operators  \cite{FeldVes}
\be \nonumber
\hat{\Delta}_\Pi =\prod_{i> j}(\Pi_i-\Pi_j) ~,~ \quad
\hat{{\Delta}}_a=\prod_{i> j}(a_i^+-a_j^+)
\ee
related, due to (\ref{con}), as 
\be
\label{identit} \hat{{\Delta}}_a=e^{ [x^2]/2} \, \hat{{\Delta}}_\Pi ~e^{-[x^2]/2}
\ee
Let us also define the bosonic-projected operators
${\tilde{\Delta}_g}$ and ${\bar{\Delta}_g}$  as
\bea
\hat{{\Delta}}_a \, \psi_B &=& {\tilde{\Delta}}_g\, \psi_B \label{Dab}\nonumber \\
\hat{{\Delta}}_\Pi \, \psi_B &=&{\bar{\Delta}}_g\, \psi_B \label{DDB}
\eea
meaning that all exchange operators $M_{ij}$ in $\hat{\Delta}_a$ and ${\hat{\Delta}_\Pi}$ are 
commuted to the right and then replaced by 1 to produce ${\tilde \Delta}_g$ and ${\bar{\Delta}}_g$ respectively.

All the above operators are antisymmetric under particle permutation, so when they act on bosonic states produce
fermionic ones. Exchange-Calogero Hamiltonians acting on $N$-body fermionic states $\psi_F$ become
\bea\label{haf}
\hat{{H}}_a  \,\psi_F&= {\tilde{H}}_{-g} \, \psi_F &= {\tilde{H}}_{g+1} \,\psi_F \\ \nonumber
\hat{{H}}_\Pi  \,\psi_F&= {\bar{H}}_{-g}\,  \psi_F &= {\bar{H}}_{g+1} \,\psi_F
\eea
From the commutativity of $\Pi_i$ it follows
\be\nonumber
\hat{H}_\Pi \,\hat{\Delta}_\Pi = \hat{\Delta}_\Pi \,\hat{H}_\Pi
\ee
Applying this relation on a bosonic state and using the projection relations (\ref{hab},\ref{DDB},\ref{haf}) we deduce
\be \label{gg+1}
{\bar H}_{g+1}\, {\bar \Delta}_g \, \psi_B = {\bar \Delta}_g \, {\bar H}_g \, \psi_B
\ee
where we used the fact that ${\bar \Delta}_g \psi_B$ is fermionic.
Since all the operators involved in (\ref{gg+1}) are local, in the sense that their action involves only the value of the
function in a small neighborhood of the $N$-dimensional point $x_1, \dots , x_N$, it follows that their
equality cannot depend on global properties of the function on which they act, such as its symmetry; therefore 
\be\nonumber
{\bar H}_{g+1}\, {\bar \Delta}_g  = {\bar \Delta}_g \, {\bar H}_g 
\ee

For $g$ integer, iteration of the above relation yields
\be\nonumber
{\bar H}_{g}\, {\bar \Delta}_{g-1} \ldots{\bar \Delta}_0 = {\bar \Delta}_{g-1}\ldots{\Delta}_0 \, {\bar H}_0
\ee
where ${\bar H}_0$ is the free $N$-body Hamiltonian.
Acting on the symmetrized free plane wave
\be\nonumber
S \, e^{i \sum_i x_iz_i} = \sum_{\pi\in S_N} e^{i \sum_i x_{\pi(i)} z_i}
\ee 
the above gives
\be \nonumber
{\bar H}_{g}\, {\bar \Delta}_{g-1}\ldots{\bar \Delta}_0 \, S \, e^{i \sum_i x_i z_i}=
\half [z^2] \, {\bar \Delta}_{g-1}\ldots{\bar \Delta}_0 \, S \, e^{i \sum_i x_iz_i}
\ee
which implies  that ${\bar \Delta}_{g-1}\ldots{\bar \Delta}_0 \, S\, e^{i \sum_i x_iz_i}$ is essentially $h_g[x,z]$ up to a proportionality constant.
To determine the constant, we observe that when all the differential operators in ${\bar \Delta}_{g}$ act on the plane wave 
$e^{i \sum_i x_iz_i}$ they produce a Vandermonde determinant in the $z_i$ variables, 
\be\nonumber
\prod_{i>j} (i z_i - i z_j ) = e^{i \pi {N(N-1)/ 4}}\, \Delta_z
\ee
This is the coefficient of the term involving no inverse powers of $x_i$, which leads to the corresponding asymptotic term
in $h_g[x,z]$. Comparing the coefficient with that in (\ref{asympt}) we obtain
\be\label{lastyes}
h_g[x,z]= (-1)^{g{N(N-1)/ 2}} \, \Delta_z^{-g} ~ {\bar \Delta}_{g-1} \ldots {\bar \Delta}_0 \, S \, e^{i \sum_i x_iz_i}
\ee
This provides the solution of the scattering Calogero model in terms of the operators ${\bar \Delta}_g$, although their complicated form does not allow for general explicit expressions.

We will now use the fact that $h_g [x,z]$ is, actually, fully symmetric under exchange of the $x_i$ and the $z_i$. This was obvious in the 2-particle case, but it is rather remarkable that it also holds in the many-body case. 
To demonstrate it, we observe that the operators ${\bar \Delta}_{0}, \dots, {\bar \Delta}_{g-1}$ appearing in (\ref{lastyes}) involve only operators of the form $\partial_{ij} = \partial_i - \partial_j$
as well as inverse powers of $x_{ij} = x_i - x_j$. All the terms are of the form
\be
{1 \over x_{ij} x_{kl} \cdots } \partial_{mn} \partial_{pq} \cdots
\label{t}
\ee
(with the indices not necessarily distinct). Since ${\bar \Delta}_{g'}$ is homogeneous with length dimension $-N(N-1)/2$, the total number of $\partial_{ij}$
operators and $x_{ij}$ terms must be $N(N-1)/2$.
The same is true for the product of operators ${\bar \Delta}_{g-1} \cdots {\bar \Delta}_0$: derivatives can be
commuted to the right to give terms of the form (\ref{t}) above, where now the total dimension is
$-gN(N-1)/2$.
These operators acting on an exponential $e^{i z_i x_i}$ will produce terms of the form
\be
z_{mn} z_{pq} \cdots z_{ij} z_{kl} \cdots {1 \over s_{ij} s_{kl} \cdots} ~,\quad s_{ij} := z_{ij} x_{ij}
\label{tz} \ee
where we introduced the dimensionless variables $s_{ij}$. The total degree of the prefactor in the $z_i$ is $gN(N-1)/2$.

Including the prefactor $\Delta_z^{-g}$ in (\ref{lastyes}), which has total degree $-gN(N-1)/2$
in the $z_i$, we deduce that all $z$-prefactors in the terms (\ref{tz}) appearing in $h_g$ are dimensionless.
Thus they must be either constants or rational expressions in the $z_{ij}$. Such rational
expressions would develop infinities for some $z_i \to z_j$ in their denominator. This is not possible, as for $x_i \to x_j$ the behavior of the
scattering solution should approach that of the corresponding 2-particle solution, which is a function of
$s_{ij} = z_{ij} x_{ij}$, and no extra poles should arise as $z_{ij} \to 0$.
We conclude that all prefactors must be constants and thus the full $h_g [x,z]$ (apart from the exponential $e^{i z_i x_i}$) is a function of the variables $s_{ij}$ alone (we have checked this fact explicitly in a few cases).
Since all these variables, as well as the exponential, are invariant under $x \leftrightarrow z$,
it follows that $h_g [x,z] = h_g [z,x]$. 
Summing over the remaining exponentials in $S e^{i z_i x_i}$ makes $h_g [z,x]$
fully permutation symmmetric.

Taking advantage of the $x \leftrightarrow z$ swap symmetry, we rewrite (\ref{lastyes}) as
\be \nonumber
h_g [x,z] = (-1)^{g{N(N-1)/ 2}} \, \Delta_x^{-g} ~ {\bar \Delta}_{g-1}[z]\ldots{\bar \Delta}_0 [z] \,S \, e^{i \sum_i x_iz_i}
\ee
where ${\bar \Delta}_{g}[z]$ are the same operators as ${\bar \Delta}_{g}$ but expressed in terms of the $z_i$ rather than
$x_i$. This is the $N$-body generalization of the relation (\ref{itero}) of the 2-body case.

The integral of interest now in (\ref{sem}) becomes, for integer $g$, 
\bea 
\nonumber &&\hskip -1cm \int {\Delta_x}^{g} \, e^{-[x^2]} \, h_g [x,z] \, [dx] \nonumber \\
\hskip 1cm&=&\int{\Delta_x}^g\, e^{-[x]^2}
(-1)^{g{N(N-1)/ 2}} \, \Delta_x ^{-g}{\bar \Delta}_{g-1}[z]\ldots{\bar \Delta}_0[z] \, S \, e^{i \sum_i x_iz_i} [dx] \nonumber\\
\nonumber &=&N! \, (-1)^{g{N(N-1)/ 2}}{\bar \Delta}_{g-1}[z]\ldots{\bar \Delta}_0[z] 
\int e^{-[x]^2+ i \sum_i x_iz_i}\, [dx]\\
&=&(-1)^{g{N(N-1)/ 2}} \, \pi^{N/ 2}\, {\bar \Delta}_{g-1}[z]\ldots{\bar \Delta}_0[z] \, e^{-{}[z^2]/4} \label{intint}
\eea
(The factor $N!$ in the third line arose from the $N!$ terms in $S \,  e^{i \sum_i x_iz_i}$, but is then reabsorbed
because we are integrating over only one of the $N!$ wedges of the full configuration space.)

It remains to determine the action of operators ${\bar \Delta}_{g}[z]$ on $e^{-[z^2]/4}$. From the
relation
\be\nonumber
\hat{{H}}_a \, a_i^+=a_i^+(\hat{{H}}_a +1)
\ee
which is a corollary of (\ref{aa+}), we deduce
\be\nonumber 
\hat{{H}}_a \hat{{\Delta}}_a =\hat{{\Delta}}_a \Bigl(\hat{{H}}_a +{N(N-1)\over2}\Bigr)
\ee
(since $\hat{{\Delta}}_a$ contains $N(N-1)/2$ creation operators).
Applying  the above intertwining relation to a bosonic state $\psi_B$ we obtain
\be \label{nice}
{\tilde{H}}_{g+1} \, {\tilde{\Delta}}_g ={\tilde{\Delta}}_g \,
\Bigl({\tilde{H}}_g+{N(N-1)\over2}\Bigr)
\ee
where we used again the projection relations (\ref{hab},\ref{Dab},\ref{haf}) and the fact that 
${\tilde \Delta}_g \psi_B$ is fermionic, as well as the locality of all operators to remove $\Psi_B$.
Applying (\ref{nice})  to the ground state harmonic Calogero wavefunction 
$\psi_0={\Delta_x}^g \, e^{-[x^2]/2}$ with energy eigenvalue $E_{g,0} = gN(N-1)/2 +N/2$ we obtain
\bea\nonumber
{\tilde{H}}_{g+1} {\tilde{\Delta}}_g \left({\Delta_x}^g \, e^{-[x^2]/2}\right) \hskip -0.2cm
& =&\hskip -0.2cm{\tilde{\Delta}}_g \left( {E_{g,0} +{N(N-1)\over2}}\right) \left({\Delta_x}^g \,
e^{-[x^2]/2}\right)\\\nonumber
&=&\hskip -0.2cm E_{g+1,0} ~{\tilde{\Delta}}_g \left({\Delta_x}^g \, e^{-[x^2]/2}\right)
\eea
This means that ${\tilde{\Delta}}_g \bigl({\Delta_x}^g \, e^{-[x^2]/2}\bigr)$ is the ground state of ${\tilde{H}}_{g+1} $. Since this ground state is unique, it follows that the two states are related as
\be\label{shift1}
{\tilde{\Delta}}_g \, \bigl({\Delta_x}^g \, e^{-[x^2]/2}\bigr) =C \, {\Delta_x}^{g+1} \, e^{-{}[x^2]/2}
\ee
with $C$ a proportionality constant. 
Similarly, applying (\ref{identit}) on a bosonic state  we get by  the same locality  argument as above
\be\label{shift2}
{\tilde{\Delta}}_g = e^{[x^2]/2}\, {\bar{\Delta}}_g \, e^{-[x^2]/2}
\ee
Combining (\ref{shift1}) and (\ref{shift2}) we obtain
\be\nonumber
{\bar \Delta}_g \left( {\Delta_x}^g \, e^{-[x^2]} \right) = C\, {\Delta_x}^{g+1} \,e^{-[x^2]}
\ee  
To find the constant $C$ it suffices to notice that the highest-power term in
${\bar{\Delta}}_g ({\Delta_x}^g \, e^{-[x^2]})$ is obtained when the highest number of derivatives in ${\bar \Delta}_g$
all act on $e^{-[x^2]}$.  
But the maximal-derivative term in ${\bar \Delta}_g$ is $\prod_{i>j}({\partial / \partial x_i}-{\partial / \partial x_j})$, and applying it to $e^{-[x^2]}$ produces $(-2)^{N(N-1)/ 2}\Delta_x e^{-[x^2]}$. So
\be
\nonumber
{\bar{\Delta}}_g\left({\Delta_x}^g \, e^{-[x^2]}\right)=(-2)^{N(N-1)/ 2}{\Delta_x}^{g+1} \, e^{-[x^2]}
\ee
Rewriting this relation in terms of $z_i /2$ instead of $x_i$ and taking into account the scaling of ${\bar \Delta}_g$
and $\Delta_x$ under rescaling of their argument we obtain
\be\nonumber
{\bar{\Delta}}_g [z] \left({\Delta_z}^g \, e^{- [z^2]/4}\right)=(-2)^{-{N(N-1)/ 2}}{\Delta_z}^{g+1} \, e^{-[z^2]/4}
\ee
Applying this relation recursively yields
\be\nonumber
 {\bar \Delta}_{g-1}[z]\ldots{\bar \Delta}_0[z] \, e^{-[z^2]/4} = 
(-2)^{-g{N(N-1)/ 2}}{\Delta_z}^{g} \, e^{-[z^2]/4}
\ee
and inserting this result in (\ref{intint}) we obtain
\be\nonumber
 \int {\Delta_x}^{g} \, e^{-[x^2]} \, h_g [x,z] \, [dx] =
2^{-g{N(N-1)/ 2}} \, \pi^{N/ 2}\,{\Delta_z}^{g} \, e^{-[z^2]/4}
\ee
This was derived for integer values of $g$. Still, it should be possible to analytically continue it to arbitrary real values, especially since the corresponding relation for $N=2$ is known to hold for arbitrary $g$.

Inserting the above expression for the integral in (\ref{sem}) we obtain our final result. In order to eliminate spurious
factors, we adopt the new rescaled definition of $\psi_\ell$
\be\nonumber
\psi_{\ell} = 2^{g{N(N-1)/ 2}} \, \pi^{-{N/ 2}}\sum_{\pi\in S_N}\prod_{i=1}^N \big(i\, a_{\pi(i)}^+\big)^{\ell_i}\, \psi_0
\ee
This absorbs all numerical factors in (\ref{sem}), which becomes
\be\nonumber
\int e^{[z^2]/4} e^{-[x^2]/2}  h_g [x,z] \, \psi_\ell [x] \, [dx]
= {\Delta_z}^{g} \, \sum_{\pi\in S_N} \prod_{i=1}^N  z_{\pi(i)}^{\ell_i} 
\ee
This is our desired mapping relation.


 \section{Conclusions and open issues} 
 We presented an explicit convolution that maps Calogero to anyon wavefunctions. Strictly speaking some steps in Section 3, relevant to the appearance of the Vandermonde prefactor, were performed by assuming integer values for the statistical parameter $g$ but are expected to hold for general $g$ by analytic continuation. This is indeed the case for $N=2$, but proving these results directly for non-integer values of $g$ and arbitrary $N$ remains an interesting open issue.
 
 Our study opens the road for
investigations of a similar mapping in related situations, in particular to spaces of nontrivial topology.
Specifically, the Calogero model can also be compactified on a periodic space, the corresponding model involving inverse-sine squared interactions and referred to as the Sutherland model. (In its basic form it does not involve one-particle potentials, although
such potentials can be added without destroying integrability \cite{polyex}.) For the Sutherland model the same question arises: does it map to
a particular projection of a 2d anyonic model? Indeed, one can view this model as a different kind of infrared regularization of the Calogero model, the long distance harmonic regulator discussed above having been traded for a compact space (the thermodynamic limit is then obtained by taking the size of the space to infinity). The corresponding 2d anyonic model should be regularized be eliminating the harmonic trap and defining it on a (perhaps partially) compact space. This could be a cylinder, which would compactify one dimension, or a 2d sphere, which would offer a fully symmetric compactification. A restriction of the angular momenta of the anyon states should also be invoked, similar to the one on the plane. This connection is currently under investigation.

Another generalization of the Calogero model involves periodization of its potential on a complex torus, leading to the 
elliptic Calogero model. A natural conjecture would be that this model would map to (particular states of) anyons on a doubly periodic space torus. Such a mapping presents several challenges. On the anyon side, FQH/anyon states on manifolds of toroidal topology are notoriously hard to write. Issues of nontrivial non-abelian representations of the global magnetic translation group (aka the ``noncommutative torus") contribute, but also the exact mathematical expressions of the states are unknown. On the Calogero side, the identification of the excitation energies and wavefunctions of the elliptic Calogero model is a hard long-standing problem, in contrast to the harmonic or Sutherland versions of the model.

What makes this problem both tantalizing and frustrating is the fact that the classical elliptic model can be obtained as a limit of an ordinary Calogero system with a large number of particles reduced by a set of discrete symmetries. Quantum mechanically, however, the imposed constraints are second class and the corresponding reduction of quantum states is not easy to obtain, notwithstanding some interesting analytical work by Langmann \cite{langmann}. Analytical work by Ruijsenaars \cite{ruijsenaars} is restricted to the simplest nontrivial case of two particles. Other connections and relations of the elliptic system to deformations of Yang-Mills theory on elliptic curves \cite{nekrasov} have also not led to explicit results.
Obtaining explicit relations between the Calogero and anyon system for the elliptic/toroidal case would contribute towards filling
this gap.

\section{Acknowledgements}
 
{The authors would like to thank Vincent Pasquier for interesting discussions and the anonymous referee for critical remarks and suggestions that helped improve the manuscript. A.P.\ acknowledges the support of NSF under grant 1519449 and the hospitality of LPTMS at Universit\'e Paris Sud (Orsay), where this work was initiated. A.P.  has also benefited from an INP CNRS  grant and from the Investissements d'Avenir LABEX PALM  grant ANR-10-LABX-0039-PALM.}

\end{document}